\documentclass[12pt,preprint]{aastex}
\usepackage{graphicx}
\begin{document}
\title{On the corona of magnetars}

\author{Yury Lyubarsky and David Eichler}

\affil{Physics Department, Ben-Gurion University of the Negev,
Beer-Sheva, Israel}

\begin{abstract}
Slow dissipation of non-potential magnetic fields in the
magnetosphere of the magnetar is assumed to accelerate particles
to hundreds MeV along the magnetic field lines.    We consider
interaction of fast particles with the surface of the magnetar. We
argue that the collisionless dissipation does not work in the
atmosphere of the neutron star because the two-stream instability
is stabilized by the inhomogeneity of the atmosphere. Rather, the
dominant dissipation mechanism is collisional Landau level
excitations followed by pair production via the deexcitation
gamma-rays ultimately leading to electrons with the energy below
the Landau energy.
We show that, because of the effects of the superstrong magnetic
field, these electrons could emerge from the surface carrying most
of the original energy so that a hot corona arises with the
temperature of $1\div 2$ MeV.
This extended
 corona is better suited than a thin atmosphere to
convert most of the primary beam energy to non-thermal radiation
and, as we show,   most of the coronal energy release is radiated
away in the hard X-ray and the soft gamma-ray bands by
Comptonization and bremsstrahlung. The radiation spectrum is a
power-law with the photon index $1<\alpha<2$. The model may
account for the persistent hard X-ray emission discovered recently
from the soft gamma-ray repeaters and anomalous X-ray pulsars and
predicts that the radiation spectrum is extended into the MeV
band.
\end{abstract}


\section{Introduction}
It is now commonly accepted that Soft Gamma Repeaters (SGRs) and
Anomalous X-ray Pulsars (AXPs) are magnetars, neutron stars with
extremely high magnetic fields, $B\sim 10^{14}\div 10^{15}$ G
(Duncan \& Thompson 1992; Paczynski 1992; Thompson \& Duncan
1995). Activity of these sources is fed by the energy of this
magnetic field (see Woods \& Thompson 2004 for a recent review).
It was found recently (Kuiper et al. 2004, 2006; Mereghetti et al.
2005, Molkov et al. 2005) that the persistent pulsed X-ray
emission from these sources has a nonthermal spectrum extending up
to $\gtrsim 100$ keV. The luminosity of this tail, $L\sim 10^{36}$
erg/s, exceeds the thermal luminosity from the star's surface. The
spectra are exceptionally hard with photon indices typically in
the range $1\div 2$ so that the luminosity peaks above 100 keV. In
this paper, we discuss origin of this emission.

According to Thompson, Lyutikov \& Kulkarni (2002), magnetic field
in magnetar's magnetosphere is generally non-potential. Slow
untwisting of the magnetic field lines generates the electric
field; particles are accelerated in this field until they produce
electron-positron pairs filling the magnetosphere and providing
charge carriers for the electric current (Beloborodov \& Thompson
2007). Pairs are formed from gamma-photons produced by cyclotron
scattering of the surface radiation of the neutron star on primary
particles. In magnetar's magnetic field, the energy of the first
Landau level is high,
\begin{equation}
\varepsilon_B=m_ec^2\sqrt{1+2B/B_q}\approx 3.5\sqrt{B_{15}}\,{\rm
MeV};\label{Landau}
\end{equation}
so that the scattered photon is immediately converted into an
electron-positron pair. Here $B=10^{15}B_{15}$ G is the magnetic
field of the star, $B_q=m_e^2c^3/\hbar e=4.4\times 10^{13}$ G. If
the surface emits radiation with the characteristic photon energy
$\varepsilon=10\varepsilon_{10}$ keV, the cyclotron scattering
occurs when the Lorentz factor of the electron reaches the value
\begin{equation}
\gamma=\frac
B{B_q}\frac{m_ec^2}{\varepsilon}=10^3\frac{B_{15}}{\varepsilon_{10}}.
\label{gamma}
\end{equation}
Note that the recoil effect should be taken into account in this
estimate (Thompson \& Beloborodov 2005). Eventually the energy is
deposited in pairs with the Lorentz factor something below
(\ref{gamma}) such that they already could not scatter the surface
photons resonantly. Nonresonant scattering is strongly suppressed
(Beloborodov \& Thompson 2007) therefore these pairs freely flow
along the magnetic field lines until they hit the surface of the
neutron star where the energy is released. The observed hard X-ray
persistent emission definitely comes from an optically thin,
tenuous plasma; this suggests that the energy is released at a
small depth, $\lesssim 1$ g/cm$^2$.

It was assumed by Thompson \& Beloborodov (2005) and by
Beloborodov \& Thompson (2007) that the two-stream instability
develops in a very narrow upper layer of the atmosphere so that
collisionless processes are responsible for deceleration of the
beam and plasma heating. The observed hard radiation was
attributed to bremsstrahlung from this atmosphere. However, the
two-stream instability is quenched when the beam is spread into a
plateau-like distribution so that only one half of the initial
beam energy is released in the optically thin domain. The beam is
in any case stopped finally at a larger depth by collisions with
the background particles. So only one half of the beam energy
could be radiated in the hard band. Of this, one half is directed
towards the star where it is mostly absorbed and reradiated back
in the soft band. One concludes that collisionless dissipation in
the atmosphere converts only 1/4 of the total energy to the hard
radiation and could not account for the fact that most of the
magnetars luminosity in the quiescent state is observed at
$\varepsilon\gtrsim 100$ keV. Moreover, as demonstrated below, the
inhomogeneity of the neutron star atmosphere stabilizes the
two-stream instability so that collisionless heating of the
atmosphere is quenched.

An alternative explanation is that the nonthermal tail is formed
via resonant scattering of the thermal surface radiation at high
altitudes, about 10 stellar radii, where the Landau energy is
already nonrelativistic. This would mean that the persistent
emission of the magnetar is associated only with a small fraction
of the magnetic field lines (those rising into the resonant
region) whereas the currents flowing along the most of magnetic
field lines and carrying most of the energy do not show up.
Moreover, careful calculations show that the resonant scattering
cannot reproduce the observed rising energy spectra of the
persistent emission (Fern\'andes \& Thompson 2007).

In this paper, we reanalyze interaction of the fast particle beam
with the surface layers of the neutron star and show that, because
of the extreme physics introduced by the ultrastrong magnetic
field,  a hot atmosphere could be formed via \emph{collisional}
processes  and these difficulties avoided. 
It will be demonstrated
below that even though these electrons appear at a relatively
large depth, they could escape upwards forming a hot layer with
the temperature $T\approx \varepsilon_B/2\approx 1\div 2$ MeV at
the top of the cold atmosphere. The reason is that the Coulomb
cross section sharply decreases when the electron energy becomes
less than $\varepsilon_B$ and the electron is restricted to move
only along the magnetic field like a bead on a wire. These
electrons could carry away a significant fraction of their initial
energy if the atmosphere is fully ionized so that there are no
ionization losses.
We show that helium atmosphere satisfies these conditions. Such an
atmosphere is expected in view of spallation of heavy nuclei
bombarded by high-energy electrons and positrons.

Bremsstrahlung radiation from this hot layer is hard enough to
populate the magnetosphere by electron-positron pairs. This pair
corona is also heated via collisionless relaxation of the primary
beam because the corona, unlike the atmosphere, is extended enough
for the two-stream instability to develop. The observed hard
X-rays are radiated via Comptonization in the corona and
bremsstrahlung in the hot atmosphere. Our model predicts a hard
radiation spectrum extending to the MeV band. In the band
$\lesssim 100$ keV, Comptonization forms a power-law spectrum with
the photon slope $1<\alpha<2$ as is observed. Because the corona
is extended, most of radiation is not intercepted by the surface
of the neutron star but rather escapes to the infinity. Therefore
the observed persistent radiation from magnetars is dominated by
hard emission.

 The paper is organized as follows. In sect.2, we analyze the two
stream instability in the neutron star atmosphere and show that
the instability is suppressed by the inhomogeneity of the
atmosphere. In sect.3, we consider deceleration of the fast
particle beam via resonant Coulomb scattering
 and development of the electron-positron cascade in the atmosphere
 (Kotov \& Kelner 1985, Beloborodov \& Thompson 2007);
 we show that the initial energy of the beam is eventually converted into
 electrons with the energy something below $\varepsilon_B$ at the depth of
 roughly 100 g/cm$^2$. In sect.4, we demonstrate that these electrons could escape
 upwards delivering most of the original energy into a hot layer at the
 top of the cold atmosphere. In sect.5, we argue that
 hard radiation from this hot layer populates the magnetosphere with
 electron-positron pairs and this pair corona emits hard radiation
 extended into the MeV band. Conclusions are presented in
 Sect.6. In Appendix A, we show that the energy the electron loses in the
 atmosphere for recoil of ions is small. In Appendix B, we
 demonstrate that helium is fully ionized in the atmospheres of magnetars.
 In Appendix C, we show that Comptonization of soft photons on mildly relativistic electrons
 in a super-strong magnetic field results in a power-law radiation spectrum with the
 photon index $\alpha\ge 1$, like in the non-magnetized case.

\section{Inefficiency of the collisionless interaction of the beam with the atmosphere}

Thompson \& Beloborodov (2005) and Beloborodov \& Thompson (2007)
suggested that the energy of the magnetospheric currents is
dissipated when the downward electron-positron beam with the
Lorentz factor (\ref{gamma}) is decelerated in a thin surface
layer of the star where two-stream instability excites Langmuir
turbulence thus providing the necessary relaxation mechanism. In
this section, we check conditions for the development of the
two-stream instability. One can conveniently express the plasma
density in the atmosphere via the Thompson optical depth, $\tau$,
as
\begin{equation}
n=\frac{\tau}{\sigma_TH};\label{n}
\end{equation}
where
\begin{equation}
H=\frac{(1+Z)T}{Am_pg}=2.6\frac{(1+Z)T_7}{Ag_{14.5}}\,\rm cm
\label{height}
\end{equation}
is the hydrostatic height of the atmosphere, $g=10^{14.5}g_{14.5}$
cm$\cdot$s$^{-2}$ the surface gravity, $A$ and $Z$ the atomic
weight and charge, correspondingly. The beam density may be
expressed via the total energy release in the magnetosphere,
$L=10^{36}L_{36}$ erg/s, as
\begin{equation}
n_b=\frac L{4\pi R_*^2
m_ec^3\gamma}=4\times10^{15}\frac{L_{36}}{\gamma_3}\,\rm
cm^{-3};\label{n_b}
\end{equation}
where $\gamma_3=\gamma/10^3$.

 The basic regimes of the two-stream
instability for a relativistic beam propagating in a
non-relativistic plasma were studied by Fainberg, Shapiro \&
Shevchenko (1970). For a monochromatic beam moving with the
Lorentz factor $\gamma$ along the strong magnetic field, the
growth rate is written in the hydrodynamic regime,
\begin{equation}
\kappa_{hydr}=\frac{\sqrt{3}}{2^{4/3}}\left(\frac{n_b}n\right)^{1/3}\frac{\omega_p}{\gamma};\label{hydr}
\end{equation}
where
\begin{equation}
\omega_p=\sqrt{\frac{4\pi e^2n}{m_e}}\label{omegap}
\end{equation}
is the plasma frequency. This formula is valid provided all the
particles in the beam are in resonance with the excited wave,
$\delta v/c<\kappa/\omega_p$. For the beam with a large energy
spread, $\delta\gamma\sim\gamma$, this condition is written as
\begin{equation}
\left(\frac{n_b}{n}\right)^{1/3}\gamma>1.\label{hydrkin}
\end{equation}
Making use of Eqs.(\ref{n}-\ref{hydr}), one can write this
condition as
\begin{equation}
\gamma>400\left(\frac{A\tau
g_{14.5}}{(1+Z)T_7L_{36}}\right)^{1/2}.\label{hydr-kin}
\end{equation}
In the opposed limit, only some fraction of the beam particles are
in resonance with the excited wave (kinetic regime). Then the
growth rate is
\begin{equation}
\kappa_{kin}\approx\frac{n_b}{n}\frac{\omega_p}{\gamma^3\delta
v^2}\approx\frac{n_b\gamma}{n}\omega_p.\label{kin}
\end{equation}
Making use of these estimates, one easily finds that $\kappa H/c$
is very large in both regimes therefore at first glance, strong
Langmuir turbulence should be excited. However, careful
consideration shows that this is not the case because at
$\kappa/\omega_p\ll 1$, the instability is easily stabilized by
plasma inhomogeneity 
(e.g. Breizman \& Ryutov 1974; Breizman 1990). Let us consider how
inhomogeneity of the neutron star atmosphere affects development
of the two-stream instability.

The instability excites the Langmuir waves with the dispersion
relation
\begin{equation}
\omega=\omega_p\left(1+\frac{3kT}{2m_ec^2}\frac{k^2c^2}{\omega_p^2}\right);\label{disp}
\end{equation}
where $T$ is the plasma temperature, $k$ the wave vector. Within
the atmosphere with the characteristic height (\ref{height}), the
plasma frequency varies with the depth, $z$, as
\begin{equation}
\frac{\delta\omega_p}{\omega_p}=\frac{\delta z}{2H}.
\end{equation}
 The frequency of the propagating wave remains constant whereas the
wave vector varies to satisfy the dispersion equation
(\ref{disp}):
\begin{equation}
\delta\omega_p+\frac{3kT}{m_ec}\delta k=0.
\end{equation}
Here we take into account that for the resonance wave,
$\omega/k\approx\omega_p/k\approx c$. The waves are amplified only
if their phase velocity is close to the beam velocity,
\begin{equation}
\vert\omega-vk\vert\lesssim\kappa.\label{res}
\end{equation}
In the hydrodynamic regime, when the beam velocity spread is less
than the width of the resonance, the condition (\ref{res}) yields
 $\delta k<\kappa_{hydr}/c$. Then one finds that
the amplification stops after the wave propagates the distance
\begin{equation}
\delta
z=\frac{6kT}{m_ec^2}\frac{\kappa_{hydr}}{\omega_p}H.\label{deltaz}
\end{equation}
A significant fraction of the beam energy could be dissipated
provided
\begin{equation}
\frac{\kappa_{hydr}\delta z}{v_g}>\Lambda;\label{instcond}
\end{equation}
where $\Lambda=10\Lambda_1$ is the logarithm of the ratio of the
beam energy to the initial energy of the Langmuir oscillation (it
is of the order of the Coulomb logarithm) and it is taken into
account that the wave propagates with the group velocity
\begin{equation}
v_g=\frac{d\omega}{dk}=\frac{3kT}{m_ec^2}c.\label{v_g}
\end{equation}
Now one finds making use of Eqs.(\ref{hydr}), (\ref{deltaz}) and
(\ref{v_g})
\begin{equation}
\frac{\kappa_{hydr}\delta z}{v_g\Lambda}=
1.3\times 10^{-6}
\frac{(1+Z)T_7^{7/6}L_{36}^{2/3}}{Ag_{14.5}\gamma_3^{8/3}\tau^{1/6}\Lambda_1}.
\label{ampl}
\end{equation}
So the condition (\ref{instcond}) is violated by a large margin.

In the kinetic regime, the condition (\ref{res}) reduces to
$\delta k/k<\delta v/c\sim (c\gamma^2)^{-1}$, which yields
\begin{equation}
\delta z=\frac{6kT}{m_ec^2\gamma^2}H.
\end{equation}
Then one finds
\begin{equation}
\frac{\kappa_{kin}\delta
z}{v_g\Lambda}=
4.3\times
10^{-4}\frac{1}{\Lambda_1\gamma_2^2}\sqrt{\frac{(1+Z)T_7}{A\tau
g_{14.5}}}.
\end{equation}
Again, the condition (\ref{instcond}) is violated by a large
margin. Therefore
the two-stream instability does not develop in the atmosphere of
the neutron star and the atmosphere could not be heated via
collisionless dissipation.

It will be shown below that a hot atmosphere could arise in the
case under consideration but as a result of \emph{collisional}
interaction of the beam with the surface layers of the star. Then
the characteristic scale of the density variation significantly
increases and moreover, the hot atmosphere emits hard X-rays
therefore the Lorentz factor of the beam decreases according to
Eq.(\ref{gamma}). Substituting in Eq.(\ref{ampl}) the
characteristic temperature of the hot atmosphere, $T=10^{10}$ K,
and the Lorentz factor $\gamma=100$, one can see that the
two-stream instability could develop in the hot atmosphere.
Therefore eventually some fraction of the beam energy could be
released via collisionless dissipation. However, at least one half
of the energy is in any case dissipated and delivered into the hot
atmosphere by collisional processes; it is these processes that
determine the temperature of the hot atmosphere.

\section{Collisional deceleration of the beam}
 Consider deceleration of a
relativistic electron-positron beam within the surface layers of
the magnetar. Let a relativistic electron move in plasma along the
strong magnetic field. If the electron energy exceeds the Landau
energy (\ref{Landau}), it efficiently scatters off ions into the
first Landau level and then immediately falls back emitting a
resonance photon; such a resonance Coulomb scattering is the main
mechanism of deceleration of a relativistic electron-positron beam
in the case under consideration (Kotov \& Kelner 1985, Beloborodov
\& Thompson 2007). The cross section for the transition of an
electron with the Lorentz factor $\gamma$ and the momentum $p$
from the ground to the $j$-th Landau level by scattering off an
ion with the charge $Ze$ is found by Bussard (1980) and Langer
(1981):
\begin{equation}
\sigma_{j0}=\frac
3{16}\frac{B_q}B\frac{Z^2\sigma_T}{(1+\gamma)^2}\frac
1{j!}\sum_{\pm}\frac{m_e^2c^2}{\mid pp_{\pm}\mid}\label{Coulomb}
\end{equation}
$$
\times
\left\{\delta_{s',-1/2}\left[(1+\gamma)^2+\frac{pp_{\pm}}{m_e^2c^2}\right]^2+
\delta_{s',1/2}\frac{2B}{B_q}\frac{p^2}{m_e^2c^2}\right\}
\epsilon_j\left(\frac{B_q}{2B}\frac{(p-p_{\pm})^2}{m_e^2c^2}\right);
$$$$
\epsilon_j(x)=\int_0^{\infty}\frac{t^jdt}{(t+x)^2}e^{-t}.
$$
Here $\sigma_T$ is the Thomson cross section, $s'$ the final spin
of electron and summation is over the final electron momentums
$p_{\pm}=\pm\sqrt{p^2-2m_ec^2jB/B_q}$. Even high energy electrons
jump predominantly on the first Landau level therefore one can
neglect excitations of higher levels. The cross section for the
transition $0\to 1$ is plotted in Fig.(1). In the limit
$mc^2\gamma\gg\varepsilon_B$, it is reduced to
\begin{equation}
\sigma_{10}=\frac{3B_q}{4B}Z^2\sigma_T\ln\left(0.413\frac{\gamma^2B_q}{B}\right).
\label{asymp}\end{equation}

The electron with the Lorentz factor $\gamma>\sqrt{2B/B_q}$ jumps
on the first Landau level having the Lorentz factor
$\gamma_1=m_ec^2\gamma/\varepsilon_B=0.15\gamma B_{15}^{-1/2}$.
The electron immediately falls back onto the background Landau
level emitting a photon with the energy $(0.5\div 1)\varepsilon_B$
as measured in the guiding center frame of the excited electron
(frame moving with the Lorentz factor $\gamma_1$); due to recoil,
the energy of the emitted photon depends on the emission angle.
After the deexcitation, the Lorentz factor of the electron remains
on average the same because in the guiding center frame, photons
are emitted forward and backward with equal probability. Thus the
electron retains only a fraction $\xi=\gamma_1/\gamma\approx
0.15B_{15}^{-1/2}$ of the total energy, most of the energy being
taken away by a photon.

\begin{figure*}
\includegraphics[width=10 cm,scale=0.8]{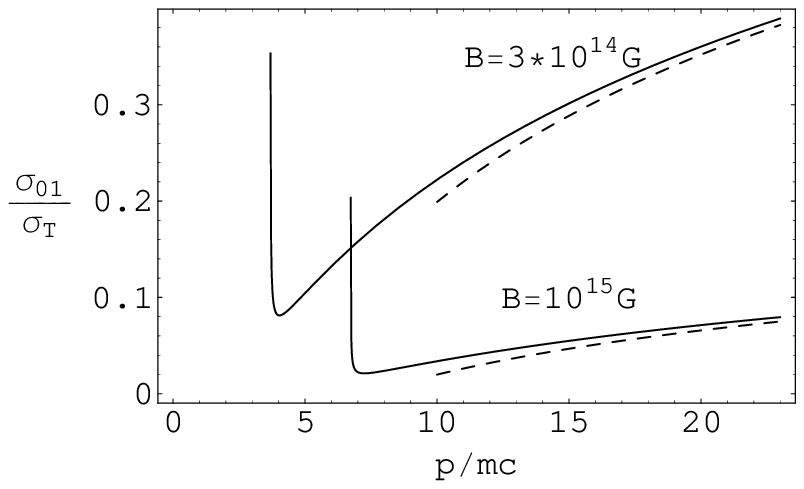}
\caption{Cross-section of the Coulomb excitation of the first
Landau level; asymptotics (\ref{asymp}) is shown by dashed lines.}
\end{figure*}

The fate of the photon depends on its polarization. Two
polarization modes could propagate in the magnetized vacuum; the
so called ordinary mode is polarized in the plane set by the
propagation direction and the background magnetic field whereas
the extraordinary mode is polarized perpendicularly to this plane.
If an ordinary photon is emitted, it is immediately converted into
an electron-positron pair provided its energy exceeds the
threshold, $\varepsilon>2m_ec^2/\sin\theta$. In magnetar's field,
this condition is satisfied for most $\theta$. In the frame of the
scattered electron, the produced electron and positron have on
average equal energies and move in opposite directions. In the
laboratory frame, most of the energy is taken by the particle
moving forward; only a fraction $1/\gamma_1^2$ of the total energy
is taken by the second particle. This means that the energy of
electrons in the beam decreases only by a fraction $\xi$ in one
scattering. If this were the only process of the beam-plasma
interaction, the energy of the particles in the beam would
decrease $\exp{(r\xi)}$ times after $r$ scattering and the total
number of scattering before the particle energy becomes less than
the Landau energy would be large,
$r\approx\xi^{-1}\ln(\gamma/\varepsilon_B)\approx 20\div 30$. The
full length of the avalanche, $l$, could be estimated by summing
the free path lengths:
$$
l=\sum_r\frac
1{\sigma_{01}n_i}\approx\frac{4B}{3B_qZ^2\sigma_Tn_i}
\int_0^r\frac{dn}{\ln(0.4\gamma^2B_q/B)-2\xi n}
$$$$
\approx \frac{2B}{3\xi
B_qZ^2\sigma_Tn_i}\ln\left[\ln\left(\frac{0.4\gamma^2B_q}B\right)\right].
$$
Here $n_i$ is the number density of ions. For $\gamma\sim 100\div
1000$, the corresponding Thomson depth is
\begin{equation}
\tau\equiv\sigma_TZn_i l\sim \frac 2Z\left(\frac
B{B_q}\right)^{3/2}=200Z^{-1}B_{15}^{3/2}.\label{tau0}
\end{equation}

An extraordinary photon does not produce pairs directly; it first
splits into ordinary photons and only then pairs may be produced.
Then the avalanche proceeds further. As the energy per particle
decreases in this case roughly twice in each step, the avalanche
penetrates the depth significantly less than that of
Eq.(\ref{tau0}) provided each emitted  O-photon is converted into
a pair. However, the energy of the photon, $(0.5\div
1)\varepsilon_B\approx 2\div 3$ MeV, is now distributed between
two photons, both or one of them could fall below the threshold
for the pair production. Then these photons either are converted
into pairs in the Coulomb field of the nucleus or experience a
Compton scattering off a background electron.

The cross section for the pair production at a nucleus is
\begin{equation}
\sigma^Z_{\pm}=\frac 7{6\pi}Z^2\alpha\sigma_T
\left[\ln\left(\frac{2\varepsilon}{m_ec^2}\right)-\frac{109}{42}\right];\label{Z}
\end{equation}
where $\alpha$ is the fine structure constant. The Compton
scattering of a photon moving along the superstrong magnetic field
was studied by Gonthier et al. (2000). In this case, the resonance
occurs only at the cyclotron fundamental
$\varepsilon=(B/B_q)m_ec^2=12B_{15}$ MeV. Above the resonance, the
scattering cross section is close to the Klein-Nishina one
\begin{equation}
\sigma_C=\frac 34\sigma_T\frac{m_ec^2}{2\varepsilon}
\left[\ln\left(\frac{2\varepsilon}{m_ec^2}\right)+\frac
12\right].\label{Compton}
\end{equation}
The pair production at nuclei dominates at energies
$\varepsilon>250Z^{-1}m_ec^2$. Note that this process comes into
play only when the photon could not be converted directly into a
pair. This occurs predominantly when E-mode resonant photons are
emitted because these photons split into O-photons with lesser
energies so that these O-photons could fall below the threshold
for the direct pair production in the magnetic field. In this
case, two pairs are produced in each step therefore the energy per
particle decreases roughly four times in each step. This means
that pairs are produced at nuclei only in the first few
generations; after this, Compton scattering dominates. The
corresponding depth varies from $\tau_Z\sim 200/Z$ at small $Z$,
when the transition energy is high and the Compton scattering
comes into play already after two generations, to $\tau_Z\sim
800/Z$, when $Z$ is large and three or four generations are
necessary.

At the Compton stage, the avalanche proceeds further. The
distribution of the scattered photons in their energies,
$\varepsilon'$, is
\begin{equation}
d\sigma_C=\frac
38\sigma_T\frac{m_ec^2d\varepsilon'}{\varepsilon^2}
\left[\frac{\varepsilon}{\varepsilon'}+\frac{\varepsilon'}{\varepsilon}+
\left(\frac{m_ec^2}{\varepsilon'}-\frac{m_ec^2}{\varepsilon}\right)^2-
2\left(\frac{m_ec^2}{\varepsilon'}-\frac{m_ec^2}{\varepsilon}\right)\right];
\end{equation}
where
$$
\frac{\varepsilon}{1+2\varepsilon/m_ec^2}\le\varepsilon'\le\varepsilon.
$$
The scattered photon takes on average a fraction
$4/[3\ln(2\varepsilon/m_ec^2)]\approx 0.2$ of the initial energy.
The photon is directed at the angle
$\sim\sqrt{4m_ec^2/\varepsilon}$ to the magnetic field therefore
it is immediately converted into a pair. The rest 0.8 fraction of
the energy is taken by the recoil electron, which emits a resonant
photon
via Coulomb scattering off an ion. If this photon is in O-mode, it
is immediately converted into a pair; if it is in E-mode, it
splits producing two photons, the energy of each of them being
roughly $(1/2)\cdot 0.8(1-\xi)\approx 1/3$ of the initial energy.
These two photons either produce pairs directly or, if they are
below the threshold for the direct pair production, experience
Compton scatterings and the process repeats again.
The Compton scattering comes into play predominantly if an E-mode
photon was emitted because a resonant O-photon is converted into a
pair immediately whereas the E-photon splits into two O-photons,
which could fall below the threshold of the direct pair
production. Therefore in the Compton channel, the energy per
particle decreases by the factor of 3 in each generation. Then the
energy per particle becomes less than $\varepsilon_B$ in a few
steps and the avalanche stops. As the Compton cross section grows
rapidly with decreasing of the photon energy and cyclotron
resonances only increase the cross section, the full length of the
avalanche is determined by the free path of a photon with the
initial energy $\varepsilon=250Z^{-1}m_ec^2$. The corresponding
depth varies from $\tau_C\sim 100$ at $Z=1$ to $\tau_C\sim 10$ at
$Z=26$.  We will argue in the next section that for our model to
be self-consistent, the avalanche should develop in the medium
composed from light elements. Spallation of nuclei by relativistic
electrons and positrons from the avalanche seems to result in
formation of a helium atmosphere; below we will consider only this
case. Then the total depth the avalanche penetrates is estimated
as $\tau=\tau_Z+\tau_C\sim 300/Z$.

In real avalanches, ordinary and extraordinary photons are emitted
alternately with approximately the same probability. As the
Coulomb cross section (\ref{Coulomb}) is significantly larger than
(\ref{Compton}) and (\ref{Z}), the longest parts of the chain is
associated with emission of extraordinary photon with subsequent
photon splitting. On the other hand, the total number of
generations is also determined by emission of extraordinary
photons because the energy per electron decreases significantly
when the photon splits. Therefore the overall length of the
avalanche is roughly the same as if only extraordinary photons
were emitted. So finally the energy of the primary beam is
converted into electron-positron pairs with the energy $\sim
\varepsilon_B/2\sim 1\div 2$ MeV at the depth of
\begin{equation}
\tau_0=\sigma_TZn_i l\sim 300/Z. \label{tau}\end{equation}

\begin{figure*}
\includegraphics[width=10 cm,scale=0.8]{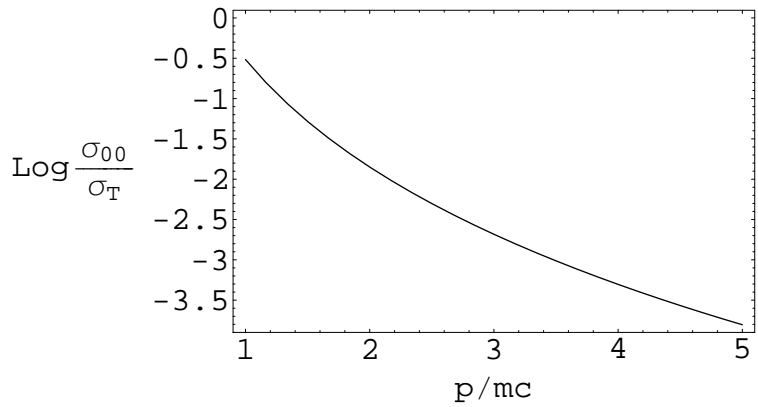}
\caption{Coulomb cross-section for the electron on the ground
Landau level, $B=10^{15}$ G.}
\end{figure*}

\section{Formation of a hot atmosphere}
Now let us consider the fate of the newly formed pairs with the
energy less than $\varepsilon_B$. Their motion is purely
one-dimensional and conservation of the energy and momentum
implies that two colliding particles of equal mass may only
exchange their energy and momenta. Therefore electron-electron
collisions do not change the state of the system and may be
neglected. Collision of a "hot" positron with a "cold" background
electron results in a "cold" positron and "hot" electron.
Collision frequency of cold particles is high therefore the cold
positron does not escape but rather eventually annihilates with
some background electron. However, the hot electron takes most of
the energy in this case  therefore most of the beam energy is
eventually stored in electrons with the characteristic energy
$\sim 1\div 2$ MeV. They diffuse within the medium colliding with
the background ions.

Collision of a one-dimensional electron with the ion may result
either to the forward scattering, after which the electron energy
remains the same to within a small recoil, or to reflection from
the ion. The cross section for the Coulomb reflection is found
from Eq.(\ref{Coulomb}) as
\begin{equation}
\sigma_{00}=\frac
34Z^2\sigma_T\frac{B_q}B\left(\frac{m_ec}p\right)^2\epsilon_0\left(\frac{2B_qp^2}{Bm_e^2c^2}\right);
\label{reflection}
\end{equation}
this cross section is plotted in Fig. 2. For
$p=0.5\varepsilon_B/c=\sqrt{B/2B_q}m_ec$ one gets
\begin{equation}
\sigma_{00}=0.6Z^2\sigma_T(B_q/B)^2=0.001Z^2\sigma_TB_{15}^{-2}.
\label{reflection1}
\end{equation}
It was shown in the previous section that the avalanche penetrates
the Thomson depth (\ref{tau}) producing electrons with the energy
$\sim\varepsilon_B/2$. One sees that for these electrons,
$\sigma_{00}n_i l=0.3B_{15}^{-2}\lesssim 1$ so they could easily
escape upwards taking away most of the energy of the avalanche.
 The cross-section of the forward scattering is larger than (\ref{reflection1})
but because recoil is small, one can neglect this process
(see Appendix A).

 The above consideration assumes that the escaping electron does
not lose energy on ionization so that the plasma in the atmosphere
is fully ionized. It is shown in the Appendix B that under the
condition of interest, helium is fully ionized. One can assume
that the layer the avalanche penetrates composed mostly of helium
because the avalanche electrons and gamma-quanta destroy heavy
nuclei(Beloborodov \& Thompson 2007). The helium is presumably
preferable to hydrogen because does not require transformation of
neutrons into protons.

One should check how much energy the electrons lose on
bremsstrahlung before they escape. In the non-magnetized medium, a
relativistic electron loses energy on bremsstrahlung after passing
the depth, in Thomson units,
\begin{equation}
\tau^{\rm br}_{0}=\frac{2\pi}{3\alpha Z(\ln
2\gamma-1/3)}.\label{br}
\end{equation}
Unfortunately bremsstrahlung in the superstrong magnetic field has
not been calculated yet however simple quasiclassical estimate
shows that the rate of the energy losses decreases in this case.

In classical electrodynamics, the energy radiated by a
relativistic electron may be written as (e.g. Landau \& Lifshitz
1995)
\begin{equation}
\delta{\cal E}
=\frac{2e^2}{3c^5}\int_{-\infty}^{\infty}\gamma^6[(\mathbf{v\cdot
w})^2+c^2w^2/\gamma^2]dt;\label{brems}
\end{equation}
where $\mathbf{w}$ is the electron acceleration.  From the
equation of motion
$$
m_e\frac{d\gamma\mathbf{v}}{dt}=e\mathbf E
$$
one finds $w_{\|}=eE_{\|}/m_ec\gamma^3$ and
$w_{\bot}=eE_{\bot}/m_ec\gamma$ where the subscripts $\|$ and
$\bot$ refer to the projections onto the direction of velocity and
onto the perpendicular direction. In the non-magnetized case, both
components of the acceleration are presented however the first
term in the square brackets in Eq.(\ref{brems}) is $\gamma^2$ less
than the second one and may be neglected. For the electron passing
the nucleus at the distance $r$ one gets assuming that the motion
is straightforward
\begin{equation}
\delta {\cal E}=\frac{\pi
Z^2e^4\gamma^2}{12r^3m_e^2c^4}.\label{brems1}
\end{equation}
Integrating the obtained relation over $r$ from
$r_{min}=\hbar\gamma/m_ec$ defined from the condition that the
energy of the emitted quanta becomes equal to the electron energy,
one gets Eq.(\ref{br}) to within a numerical factor and the
logarithmic term.

In the superstrong magnetic field, the electrons move only along
the field therefore only the first term in the square brackets in
Eq.(\ref{brems}) remains. Simple calculation shows that in this
case $\delta {\cal E}$ is $3\gamma^2$ less than that of
Eq.(\ref{brems1}). One can expect that the full bremsstrahlung
cross section in the superstrong magnetic field decreases by the
same factor. Of course quantum treatment of the process is
necessary to rigorously justify this conclusion, but assuming that
the above consideration is basically correct, the electron in the
superstrong magnetic field loses the energy on bremsstrahlung
after passing the depth $\tau^{\rm br}=3\gamma^2\tau^{\rm
br}_{0}$. For electrons with the energy $\sim\varepsilon_B/2$, one
gets $\tau^{\rm br}=6000Z^{-1}B_{15}$, which is significantly
larger than the depth (\ref{tau}) the avalanche penetrates. This
means that one can neglect bremsstrahlung energy losses.

So hot electrons freely escape upwards taking away most of the
beam energy. Charge neutrality implies that the necessary amount
of background ions rises together with the electrons so that a hot
atmosphere with the temperature $T\approx \varepsilon_B/2\approx
1\div 2$ MeV and the characteristic height
$H=26(1+Z)T_{10}/Ag_{14.5}$ m is formed.

\section{Pair corona}

The hot atmosphere is cooled via bremsstrahlung; radiation is
dominated by photons with the energy $\sim T\sim 1\div 2$ MeV. One
half of this radiation is directed towards the star; it is
absorbed and eventually is reradiated as thermal emission. The
other half of radiation is radiated away. In the magnetosphere,
hard photons could be converted into electron-positron pairs. The
pair production rate may be written as $\dot N=\zeta L/m_ec^2$
where the numerical factor $\zeta<1/2$ takes into account
uncertainties in the radiation spectrum and the field geometry.
Within the light travel time, $R_*/c$, the magnetosphere will be
filled by pairs with the density $n=8\times 10^{18}\zeta L_{36}$
cm$^{-3}$. The characteristic inhomogeneity scale of the corona is
of the order of the star radius, which is large enough for the
two-stream instability to develop in the corona. The condition for
the instability (\ref{instcond}) is satisfied provided the
left-hand side of Eq.(\ref{ampl}) exceeds unity. Substituting in
this relation $H$ by $R_*$ and making use of Eqs.(\ref{hydr}),
(\ref{omegap}) and the above estimate for the density of the
corona, one gets
$$
\frac{\kappa_{kin}\delta
z}{v_g\Lambda}=10^3\frac{L_{36}^{1/2}}{\zeta^{1/6}\gamma^{8/3}_3\Lambda_1}.
$$
Note that the condition (\ref{hydrkin}) is satisfied in the corona
at any reasonable parameters therefore the instability is
hydrodynamic. So after the corona is formed, the primary beam
would experience collisionless relaxation. Then about one half of
the beam energy is spent on heating of the corona whereas the
other half will be delivered to the surface of the star where the
the electron-positron avalanche is developed and the energy of the
beam is eventually transferred to electrons with the energy
$\sim\varepsilon_B/2\sim 1\div 2$ MeV each. It was argued in the
preovious sections, that these electrons escape upwards delivering
most of their energy into the tenious, hot atmosphere.

The corona is efficiently cooled by Comptonization of the thermal
emission from the surface. As the energy of the photons is well
below the Landau energy, only O-mode radiation is scattered
(polarized in the plane set by the direction of propagation and
the magnetic field). Thermal energy stored in the star interior is
transferred to the surface and radiated away by E-mode photons
because their opacities are $(Bmc^2/B_q\varepsilon)^2>>1$ times
less than those for O-mode photons. Soft O-mode photons could be
emitted only if surface layers of the star are heated. As hard
radiation from the hot atmosphere illuminates the underlaying cold
atmosphere, some fraction of this radiation will be absorbed and
reradiated in the soft band. It is this thermal O-mode radiation
that could be a soft photon source for Comptonization.

The scattering cross section for the O-mode photons is
$\sigma=\sigma_T\sin^2\theta'$ where $\theta'$ is the angle
between the propagation direction and the magnetic field in the
proper electron frame. The relativistic electron "sees" radiation
at the angle $\theta'\sim 1/\gamma$ therefore the cooling rate
decreases $\gamma^2$ times as compared with the non-magnetized
case and may be estimated as (Beloborodov \& Thompson 2007)
\begin{equation}
\left(\frac{d{\cal E}}{dt}\right)_C=-\sigma_TUc;\label{compcool}
\end{equation}
where $\cal E$ is the energy of the electron (assumed to be larger
than $m_ec^2$), $U$ the radiation energy density. Writing the
radiation energy density as $U=L/4\pi R_*^2c$, one finds that the
electron cooling time
\begin{equation}
t=4\times 10^{-5} {\cal E}_{\rm MeV}L^{-1}_{36}
\end{equation}
is comparable with the light travel time.

The observed radiation is a superposition of the bremsstrahlung
radiation from the hot atmosphere and the Comptonization radiation
from the corona; the radiation spectrum extends to the
characteristic particle energy of $1\div 2$ MeV. Bremsstrahlung
has a flat intensity spectrum (the photon index $1$) at
$\varepsilon\ll T$ whereas unsaturated Comptonization produces a
power-law spectrum with the photon spectral index $\alpha>1$
depending on the parameters of the system (see Appendix C). As the
luminosity of the corona is larger than the thermal luminosity of
the surface, which provides soft photons for Comptonization, the
slope should be hard enough $\alpha<2$. Therefore the
low-frequency part of the spectrum is dominated by Comptonization
and could exhibit a variety of spectral indices in the range
$1<\alpha<2$ as is observed. The observed high pulsed fraction is
naturally explained by the fact that the scattering cross section
for the ordinary mode photons is highly anisotropic.


When the pairs fill the magnetosphere, the energy release stops
because the pair plasma shorts out the induction electric field.
Energetic primary particles, which have already filled the
magnetosphere, disappear at the star's surface for about the light
travel time. At the next stage, the hot atmosphere and the corona
are cooled. The cooling time is of the order of the light travel
time. An important point that positrons could not survive for the
larger time because they are annihilated when hitting the surface
of the star. When a positron falls from the corona onto the
surface, it exchanges energy with an electron in the cold
atmosphere and annihilates\footnote{Note that even though the
annihilating particles are cold, the annihilation line is not
formed. In the superstrong magnetic field, only longitudinal
(along the field) component of the momentum is conserved therefore
the two-photon annihilation results in photons with generally
different energies, the annihilation spectrum being extended from
0 to 1 MeV (Kaminker, Pavlov \& Mamradze 1987).}.

The corona is expected to be extremely unsteady. Even though most
of the energy is taken away by the electron, which reflects from
an ion and goes upwards, this electron cannot rise directly into
the corona because of the charge neutrality and would remain in
the hot atmosphere. The corona is replenished by the hydrodynamic
expansion of this hot atmosphere, and  cooled by Compton losses.
The Compton losses may or may not be compensated by collisionless
interaction with the primary beam. Therefore the corona may be
depleted by Compton cooling; in any case, we expect that half the
primary beam energy makes it down to the hot atmosphere, so that
new matter is constantly being added to the corona, and  this very
likely leads to a non-steady situation. Because the coronal matter
can propagate the currents, the displacement currents and
attendant primary beam  are switched off, while Compton cooling
continues. When the temperature of the hot atmosphere falls below
$m_ec^2$, the pair production stops and the plasma density in the
corona falls below the critical value necessary to maintain the
magnetospheric currents. Then the displacement current arises
again and the next cycle of energy release starts. So one can
expect strong fluctuations of the radiation at the timescale of
$\Delta t \ge 10^{-4}$ s. The possibility exists of revealing them
by analysis of  the photon statistics. If a source emits radiation
in separate bursts of the characteristic duration $\Delta t \equiv
10^{-3}\Delta t_{-3}$, the probability distribution for a pair of
photons to be detected within the time interval $t$ should differ
significantly from a Poisson distribution at $t\sim \Delta t$. For
a collecting area of $10^3A_3$cm$^2$, and a photon flux of
$10^{-3}F_{-3}$cm$^{-2}$, an observation interval of $10^5t_5$s
should contain $10^5A_3F_{-3}t_5$ photons and, for Poisson arrival
statistics, $10^2A_3F_{-3}^2t_5\Delta t_{-3}$ pairs of photons
arriving within $\Delta t$ of each other. The non-steady nature of
the hard coronal emission (this emission in fact  dominates the
thermal surface emission already at a few keV), which produces
deviations from Poisson arrival statistics, is therefore
detectable with  a sufficiently powerful detector and large
exposure times.

\section{Conclusions}
We considered dissipation of the energy released in the
magnetosphere of the magnetar in the course of slow relaxation of
non-potential magnetic fields. A basic physical picture was
proposed by Thompson et al. (2002) and recently elaborated by
Thompson \& Beloborodov (2005) and Beloborodov \& Thompson (2007).
When the plasma density in the magnetosphere falls below a
critical value necessary to maintain the magnetospheric currents,
an induction electric field arises and initiates an
electron-positron avalanche resembling that in pulsars. The fast
electron-positron flow hits the surface of the star where the
released energy is dissipated. The observed very hard spectra of
the persistent X-ray emission from SGRs and AXPs (Kuiper et al.
2004, 2006; Mereghetti et al. 2005, Molkov et al. 2005) imply that
most of the energy is released in a very hot and tenuous plasma.
It was assumed by Thompson \& Beloborodov (2005) and Beloborodov
\& Thompson (2007) that the flow loses a significant fraction of
its energy in a thin surface layer where the two-stream
instability develops so that the plasma is heated by collisionless
processes. Collisionless heating is balanced by bremsstrahlung
radiation and the equilibrium temperature about 100 keV is
achieved.

We reanalyse interaction of the fast plasma flow with the surface
of the magnetar and conclude that this mechanism is incapable of
heating the atmosphere because strong density gradient in the
atmosphere of the neutron star suppresses the two-stream
instability.  We propose, rather, that a hot, tenuous
atmosphere/corona could arise due to specific properties of
Coulomb scattering in the superstrong magnetic field, mainly due
to one-dimensional character of the electron motion. Within the
upper layers of the neutron star, the flow energy is transferred,
via an electron-positron avalanche, to electrons with the energy
something less than the Landau energy ($\sim 1\div 2 $ MeV in the
magnetar's magnetic field). This occurs at a significant depth,
$\sim 100$ g/cm$^2$, however these electrons are not thermalized
but rather escape upwards taking away most of the initial flow
energy.  The reason is that collisions between electrons in
one-dimension do not result in relaxation; after the collision,
the two energies of the two electrons are the same as the initial
energies. On the other hand, collisions with ions result only in
nearly elastic reflection.
These electrons form a hot atmosphere (the necessary amount of
ions accompany the electrons in order to maintain charge
neutrality) with the temperature $\sim 1\div 2$ MeV, which is an
order of magnitude larger than in the model by Thomson \&
Beloborodov (2005) and Beloborodov \& Thompson (2007). Hard
radiation from this atmosphere is generated via bremsstrahlung.
Pairs are easily produced in this radiation field; they fill the
whole magnetosphere forming a hot corona. Collisionless
interaction of the primary beam with the pair plasma in the corona
heats the pairs even more; they are cooled by Comptonization so
that the overall spectrum of the source is a superposition of the
bremsstrahlung radiation from the hot atmosphere and a
Comptonization radiation from the corona.

The extended corona radiates in all directions so that only a
small, $<1/2$, fraction of the radiated energy is intercepted by
the surface of the star. Therefore the observed luminosity is
dominated by the hard radiation. The spectrum is extended to MeV
band. Unsaturated Comptonization generates a power-law spectrum,
which is generally steeper than the flat bremsstrahlung spectrum
therefore radiation from the corona dominates in the range
$\lesssim 100$ keV. The photon spectral slope is $1<\alpha<2$, as
is observed. The energy release in the magnetosphere occurs
spasmodically at the time scale of at least the light travel time:
the pairs short out the induction electric field in the corona and
the energy release stops until the corona cools down, then the
charge starvation necessitates the displacement current and the
process starts again. Therefore one can expect strong fluctuations
of the radiation at the time scale of $\ge 10^{-4}$s.

We are grateful to Rashid Shaisultanov for help. Y.L acknowledges
support from the German-Israeli Foundation for Scientific Research
and Development. D.E. acknowledges support from the United States
- Israel Binational Science Foundation and from an Israel Science
Foundation Center of Excellence Grant.

\section*{Appendix A. Electron-ion collisions: recoil effect}
When the electron with the energy less than $\varepsilon_B$ passes
an ion, the energy is transferred only due to recoil effect. In
magnetar's field, most of the ions are in the ground Landau state
and a scattering occurs if the ion makes a transition to the first
level with the energy
$$
\varepsilon_{Bi}=\frac{ZeB}{m_ic}=5\frac ZAB_{15}\, {\rm keV};
$$
where $m_i=Am_p$ is the ion mass. The cross section for
collisional transitions between the ion Landau levels was found by
Langer (1981); for the transition between the ground and the first
levels it is written as
$$
\sigma=\frac{3B_q}{8B}Z\sigma_T\frac{m_e^2c^2\gamma\gamma'}{pp'}(\ln\Lambda-0.577);
$$
where
$$
\Lambda^{-1}=\frac{B_q}{B}\left[\left(\frac{p-p'}{m_ec}\right)^2-(\gamma-\gamma')^2\right].
$$
The conservation laws imply
$$
m_ec\gamma+m_ic=m_ec\gamma'+\sqrt{m_i^2c^2+(p-p')^2+2\varepsilon_{Bi}m_i};
$$
which yields
$$
 \gamma-\gamma'=\varepsilon_{Bi};
$$
and
$$
\Lambda^{-1}=\frac{\varepsilon_{Bi}}{m_ic^2}\left(\frac{m_ec\gamma}p-\frac
14\right).
$$
As the electron looses only a small fraction of the energy in a
scattering, $\gamma-\gamma'\ll\gamma$, one can conveniently define
the effective cross-section for the energy loss
$$
\tilde{\sigma}\equiv [(\gamma-\gamma')/\gamma]\sigma=2\times
10^{-3}\frac{Z^2}{A\gamma}[1+0.08\ln(A^2/ZB_{15})]\sigma_T.
$$
It was shown in Sect. 3 that the electron-positron avalanche stops
at the depth (\ref{tau}) forming eventually a cloud of electrons
with the energy $\sim\varepsilon_B/2$. The electron with this
energy loses the fraction of its energy
$$
\frac{\Delta E}E=\tilde{\sigma}n_il=0.1\frac
1{AB^{1/2}_{15}}[1+0.08\ln(A^2/ZB_{15})]
$$
when rising from the depth (\ref{tau}). This estimate assumes that
the electron moves straightforwardly but not diffuses upwards.
This is justified because the layer of the depth (\ref{tau}) is
transparent for the Coulomb reflection (see the estimate
(\ref{reflection1})). So the fraction of the energy the electron
loses for recoil process is small.

\section*{Appendix B. Ionization equilibrium.}

Ionization equilibrium in the super-strong magnetic field is still
a subject of intense research, see recent reviews by Lai (2001)
and Harding \& Lai (2006) and references therein. Here we present
rough estimates for the helium plasma. The ionization energy of
the hydrogen-like ion is
$$
Q=0.16Z^2\left[\ln\frac{\hbar^3B}{m_e^2e^3cZ^2}\right]^2 {\rm
a.u.}
$$
For the helium ion, one gets $Q=2.39(1+0.086\ln B_{15})$ keV. When
the atom moves, the ionization energy decreases so that the above
value gives the estimate of the ionization temperature from above.

The ionization equilibrium He$^{++}\rightleftarrows$He$^++e$ is
written as
$$
\frac{n_en_{++}}{n_+}=\frac{Z_eZ_{++}}{Z_+};
$$
where $n_e$, $n_{++}$ and $n_+$ are the number densities of free
electrons, He$^{++}$ and He$^+$, correspondingly, $Z_e$, $Z_{++}$,
and $Z_+$ their partition functions. The partition function of the
strongly magnetized electrons is
$$
Z_e=\frac{eB}{2\pi\hbar
c}\left(\frac{m_eT}{2\pi\hbar^2}\right)^{1/2}. \eqno(B1)
$$
The ratio of partition functions of He$^{++}$ and He$^+$ is
dominated by the ionization energy and may be presented as
$$
\frac{Z_{++}}{Z_+}\approx\exp\left(-\frac QT\right).
$$
Now the temperature of ionization (when $n_{++}=n_+$) is found as
$$
T_{ion}=Q\left\{\ln\left[\frac{3eB}{2\pi n\hbar
c}\left(\frac{m_eT}{2\pi\hbar^2}\right)^{1/2}\right]\right\}^{-1}=
1.7\times 10^6\frac{(1+0.09\ln
B_{15})^2}{1+0.06\ln\left(B_{15}/\tau g_{14.5}\right)} {\rm K}.
$$
This means that helium is fully ionized in magnetar's atmosphere
with the temperature $T=0.5\div 1$ keV.

\section*{Appendix C. Comptonization in a superstrong magnetic field}

 Here we demonstrate that Comptonization of soft photons in the
superstrong magnetic field results in a power-law spectrum with
the photon index $\alpha\ge 1$, like in the non-magnetized case.
Compton scattering of soft photons on hot electrons results in a
photon flux over the spectrum from the initial energy,
$\varepsilon_0$, to the spectral region $\varepsilon\sim T$. If
the optical depth of the source is large enough, the photons are
accumulated at $\varepsilon\sim T$ and the equilibrium
Bose-Einstein spectrum, $N_{\rm
BE}=\{\exp[(\eta+\varepsilon)/T]-1\}^{-1}$ is formed.
This regime is called saturated Comptonization.
In the medium of the moderate optical depth, photons are not
accumulated but rather escape and therefore a power law spectrum
could be formed in the range $\varepsilon_0\ll\varepsilon\ll T$.
Here we show that the same occurs also in the magnetic
field so strong that the
electrons populate only the ground Landau level. For
nonrelativistic temperatures, Comptonization in the superstrong
magnetic field was studied by Lyubarskii (1987a,b) in the
Focker-Plank approximation. Here we allow relativistic
temperatures but restrict ourselves only to low photon energies
when one can neglect recoil.

Let us first consider scattering on electrons moving along the
magnetic field with some momentum $p$; the number density of this
electrons is $f(p)dp$; where $f(p)$ is the electron distribution
function. The kinetic equation for photons is easily written in
the proper frame of these electrons as
$$
\left(\frac{\partial}{\partial t'}+c\mathbf{l'}\frac{\partial
}{\partial \mathbf{r}'}\right)
 n'(\mathbf{r}',\varepsilon',\mathbf{l}')=cf'dp'\int
d\varepsilon'_1d\Omega'_1\delta(\varepsilon'_1-\varepsilon')
\frac{\partial\sigma}{\partial \Omega'}
[n(\mathbf{r}',\varepsilon'_1,\mathbf{l}'_1)-n(\mathbf{r}',\varepsilon',\mathbf{l}')];
$$
where $\mathbf{l}$ is the direction of propagation of photons,
$n(\mathbf{r},\varepsilon,\mathbf{l})$ the phase density of
photons and prime marks quantities measured in the proper electron
frame. The scattering cross-section is
$$
\frac{\partial\sigma}{\partial \Omega'}=\frac
3{8\pi}\sigma_T\sin^2\theta'\sin^2\theta'_1;
$$
where $\theta$ and $\theta_1$ is the angles between the photon
direction and the magnetic field before and after the scattering,
correspondingly. One can transform this equation to the laboratory
frame taking into account that the distribution functions are
relativistic invariants, $f'(p')=f(p)$,
$n'(\varepsilon',\mathbf{l}')=n(\varepsilon,\mathbf{l})$, as well
as the expressions $\varepsilon d\varepsilon d\Omega$ and
$\varepsilon (\frac{\partial}{\partial
t}+\mathbf{l}\frac{\partial}{\partial \mathbf{r}})$. Summation
over all electrons yields
$$
\frac{\partial n}{\partial t}+c\mathbf{l}\frac{\partial
n}{\partial \mathbf{r}}=\frac 3{8\pi}c^6\sigma_T\int
[n(\mathbf{r},\varepsilon_1,\mathbf{l}_1)-n(\mathbf{r},\varepsilon,\mathbf{l})]
$$$$
\times\delta[\varepsilon(c-v\cos\theta)-\varepsilon_1(c-v\cos\theta_1)]
\frac{\sin^2\theta\sin^2\theta_1 f(p)dpd\varepsilon_1d\Omega_1}
{\gamma^6(c-v\cos\theta)(c-v\cos\theta_1)^3}.
$$
This equation describes Comptonization of photons with the energy
larger than the energy of the seed photons, $\varepsilon_0$, but
small enough for the recoil effect to be neglected,
$\varepsilon_0\ll\varepsilon\ll \min(T,m_ec^2/T)$.

In the steady state case, $\partial/\partial t=0$, solution to
this equation has a power law form
$$
n(\mathbf{r},\varepsilon,\mathbf{l})=J(\mathbf{r},\mathbf{l})\varepsilon^{-(2+\alpha)};
$$
where the spatial function $J$ satisfies the equation
$$
\mathbf{l}\frac{\partial J(\mathbf{r},\mathbf{l})}{\partial
\mathbf{r}}=\frac 3{8\pi}c^5\sigma_T\int
\left[\left(\frac{c-v\cos\theta_1}{c-v\cos\theta}\right)^{2+\alpha}J(\mathbf{r},\mathbf{l}_1)-
J(\mathbf{r},\mathbf{l})\right]
\frac{\sin^2\theta\sin^2\theta_1f(p)dpd\Omega_1}
{\gamma^6(c-v\cos\theta)(c-v\cos\theta_1)^4}.\eqno(B1)
$$
The photon power index, $\alpha$, could be found as an eigenvalue
of this equation. It is determined by the electron distribution
function and by the geometry and the optical depth of the source.
It is beyond the scope of the present paper to solve this equation
(solution of a similar problem for a nonmagnetized plasma is given
by Titarchuk \& Lyubarskij (1995)). Let us only demonstrate that
one can expect $\alpha>1$.

In the infinite homogeneous medium, the left-hand side of Eq.(B1)
is zero therefore one gets the equation
$$
\int \left[(c-v\mu_1)^{2+\alpha}J(\mu_1)-
(c-v\mu)^{2+\alpha}J(\mu)\right]
\frac{(1-\mu^2)(1-\mu^2_1)f(p)dpd\mu_1}
{\gamma^6(c-v\mu)^{3+\alpha}(c-v\mu_1)^4}=0;\eqno(B2)
$$
where $\mu=\cos\theta$. This equation has an evident solution
$\alpha=-2$, $J(\mu)=\it const$, which is nothing more than the
low frequency part of the equilibrium Bose-Einstein spectrum. We
are interested in a solution with a non-zero photon flux over the
spectrum; in the infinite medium, such a solution implies that
except of the soft photon source at $\varepsilon=\varepsilon_0$,
there is a sink at some large enough energy $\varepsilon_{\rm
sink}$; then Eq.(B2) describes the region
$\varepsilon_0<\varepsilon<\varepsilon_{\rm sink}$. In a
non-magnetized plasma, the solution with a non-zero photon flux
over the spectrum is $n\propto \varepsilon ^{-3}$ (Kats,
Kontorovich \& Kochanov 1978) so that the intensity spectrum is
flat. In order to see that the same spectrum ($\alpha=1$ in our
notations) satisfies also Eq.(B2) note that at $\alpha=1$, the
integrand in the left-hand side of this equation is antisymmetric
with respect to exchange $\mu\leftrightarrow\mu_1$. Therefore the
integral from the left-hand side of Eq.(B2) over $\mu$ vanishes
identically. This means that a finite linear homogeneous set of
equations, which could be obtained from Eq.(B2) by discrete
approximation of the integral, is linearly dependent and therefore
it has a nontrivial solution. ({\it The formal proof}. Denote the
linear operator in the left-hand side of Eq. (B2) at $\alpha=1$ as
$\cal L$ and introduce the standard notation for the scalar
product of functions $(\psi,\phi)\equiv\int_{-1}^1\psi\phi d\mu$.
Now one can write that $(e,{\cal L}J)=0$ for $e=const$ and an
arbitrary $J$. Then $({\cal L}^*e,J)=0$ so that $e=const$ is a
nontrivial solution to the conjugate equation ${\cal L}^*e=0$. In
this case, the equation ${\cal L}J=0$ also has by the Fredholm
alternative a nontrivial solution.)

Thus the spectrum with the slope unity is formed in the infinite
medium; then all photons produced at $\varepsilon_0$ reach
$\varepsilon_{\rm sink}$. In the case of a finite optical depth,
the photons escape so that the spectral photon density should
decrease with the frequency faster than in the infinite medium.
Therefore unsaturated Comptonization in the super-strong magnetic
field generates power law spectra with the slope $\alpha>1$ like
in the non-magnetized plasma.

\end{document}